\makeatletter
\@ifundefined{@parse@version@dash}{%
\def\@parse@version#1{\@parse@version@0#1}
\def\@parse@version@#1/#2/#3#4#5\@nil{%
\@parse@version@dash#1-#2-#3#4\@nil}
\def\@parse@version@dash#1-#2-#3#4#5\@nil{%
  \if\relax#2\relax\else#1\fi#2#3#4 }
}{}
\makeatother  

\documentclass[prl,twocolumn,superscriptaddress,showpacs,floatfix,longbibliography]{revtex4-2}
\usepackage{mathrsfs,braket}
\usepackage{amssymb, amsbsy, amsmath, latexsym, dsfont, array, layout, graphicx,mathrsfs,color,ulem,bm}
\usepackage[colorlinks=true,citecolor=blue,urlcolor=blue,linkcolor=blue]{hyperref}

\begin{document}
\title{Aharonov-Bohm Caging and Inverse Anderson transition in Ultracold Atoms}

\author{Hang Li}
\thanks{These authors contributed equally to this work}
\author{Zhaoli Dong}
\thanks{These authors contributed equally to this work}
\affiliation{%
Interdisciplinary Center of Quantum Information, State Key Laboratory of Modern Optical Instrumentation, Zhejiang Province Key Laboratory of Quantum Technology and Device, Department of Physics, Zhejiang University, Hangzhou 310027, China
}%
\author{Stefano Longhi}
\affiliation{%
Dipartimento di Fisica, Politecnico di Milano, Piazza L. da Vinci 32, I-20133 Milano, Italy
}%
\affiliation{%
IFISC (UIB-CSIC), Instituto de Fisica Interdisciplinar y Sistemas Complejos, Palma de Mallorca, Spain
}%
\author{Qian Liang}
\affiliation{%
Interdisciplinary Center of Quantum Information, State Key Laboratory of Modern Optical Instrumentation, Zhejiang Province Key Laboratory of Quantum Technology and Device, Department of Physics, Zhejiang University, Hangzhou 310027, China
}%
\author{Dizhou Xie}
\thanks{%
Current address: Institut f$\ddot{u}$r Experimentalphysik und Zentrum f$\ddot{u}$r Quantenphysik, Universit$\ddot{a}$t Innsbruck, 6020 Innsbruck, Austria
}
\affiliation{%
Interdisciplinary Center of Quantum Information, State Key Laboratory of Modern Optical Instrumentation, Zhejiang Province Key Laboratory of Quantum Technology and Device, Department of Physics, Zhejiang University, Hangzhou 310027, China
}%

\author{Bo Yan}
\email{yanbohang@zju.edu.cn}
\affiliation{%
Interdisciplinary Center of Quantum Information, State Key Laboratory of Modern Optical Instrumentation, Zhejiang Province Key Laboratory of Quantum Technology and Device, Department of Physics, Zhejiang University, Hangzhou 310027, China
}%

\date{\today}

\begin{abstract}
Aharonov-Bohm (AB) caging, a special flat-band localization mechanism, has spurred great interest in different areas of physics. AB caging can be harnessed to explore the rich and exotic physics of quantum transport in flatband systems, where geometric frustration, disorder and correlations act in a synergetic and distinct way than in ordinary dispersive band systems. In contrast to the ordinary Anderson localization, where disorder induces localization and prevents transport, in flat band systems disorder can induce mobility, a phenomenon dubbed inverse Anderson transition. Here, we report on the experimental realization of the AB cage using a synthehtic lattice in the momentum space of ultracold atoms with tailored gauge fields, demonstrate the geometric localization due to the flat band and the inverse Anderson transition when correlated binary disorder is added to the system. Our experimental platform in a many-body environment provides a fashiinating quantum simulator where the interplay between engineered gauge fields, localization, and topological properties of flat band systems can be finely explored.
\end{abstract}

\maketitle

\begin{figure*}[!t]
	\centering
	\includegraphics[width=0.75\linewidth]{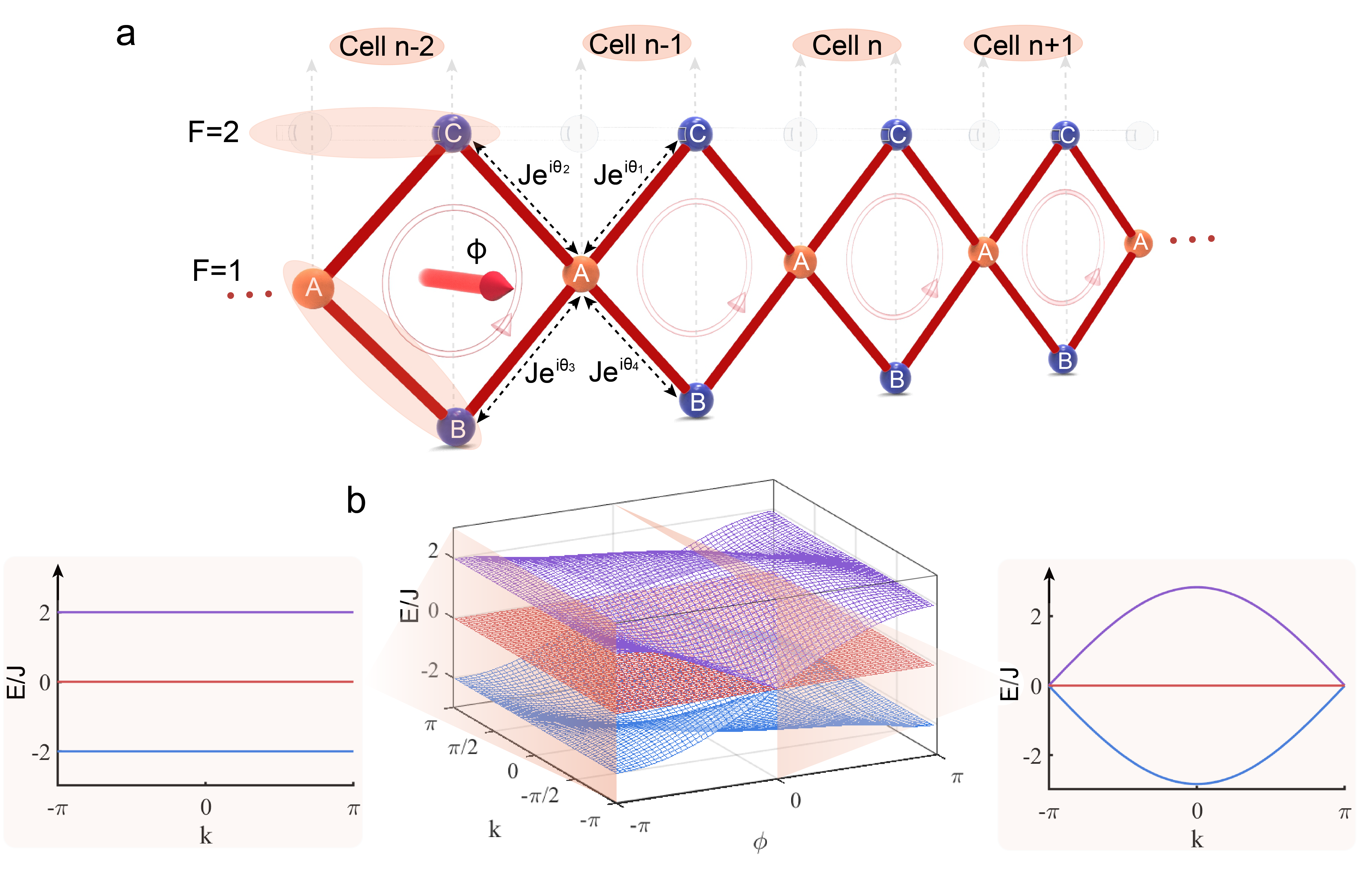}
\caption{\textbf{The experimental scheme and energy spectrum of Aharonov-Bohm cage.} \textbf{a.} Schematic diagram of a rhombic  lattice with three sites ($A_n$, $B_n$ and $C_n$) per unit cell. The sites $A_{n} (n=0,1,...)$ are put on the $F=2$ energy level, while the $B_{n}$ and $C_{n}(n=0,1,...)$ sites are rooted in the $F=1$ energy level. The complex hopping coefficients between nearest-neighboring sites are denoted as $J e^{i\theta_{m}} (m=1,2,3,4)$.  \textbf{b.} The energy dispersion $E(k)$ of the chain as a function of the flux $\phi$ (middle). The left and right insets display the dispersion relation for $\phi=\pi$ and $\phi=0$, respectively.}
	\label{f2}
\end{figure*}

{\bf Introduction.}
Exploring the deep insights into localization, disorder and transport is of upmost relevance in different areas of physics and in modern quantum technologies, ranging from condensed matter physics~\cite{Anderson,2010_RMPTI1,2011_RMPTI2} to a variety of artificial quantum and classical systems~\cite{2012_Supercon,2012_ions,2012_qugas,2008_RMPcold,2019_RMP1,2020_Rydberg,2019_RMP2,2013_NVreview,Segev2013}. When sufficiently strong static uncorrelated disorder is added to a regular lattice structure, complete localization of wave functions and the absence of transport is usually observed for non-interacting particles as a result of Anderson localization~\cite{Anderson}.  A different approach to create localization is inducing a flat-band spectrum: here destructive interference among different propagation paths is realized via the special lattice geometry, with the formation of perfectly localized compact modes \cite{FlachReview2018}. The Aharonov-Bohm (AB) caging \cite{Mosseri1998} provides a paradigmatic example  of the latter type of localization, induced by an artificial gauge field. With the development of modern quantum technologies, a variety of platforms, mainly based on ultracold quantum gas~\cite{2009_Nature,2013_prl,2013_prl2,2014_NPhy,2020_prl}, superconducting circuits~\cite{1999_prl,2017_prx} and photonic crystals~\cite{2014_NPr,2017_NPho,2019_RMP2} can be applied for creating artificial synthetic magnetic flux, thus studying the flat-band localization. Recently, the flat-band localization in the AB cage has been demonstrated in photonic lattice systems with curved or auxiliary waveguides~\cite{ABcage_prl, ABcage_nc}. 
Owing to the diverging effective mass in a flat band, the system becomes very sensitive to disorder and interactions,
and the emerging phenomena, such as  criticalilty and multifractality, can significantly deviate from
conventional Anderson localization~\cite{2001_prb,Inverse_prl,2010_prb,2014_prl,2014_epl,2017_epjb,2018_prb,2018_prb2,Baboux2016,Liu2022}.
An important example is the inverse Anderson transition, i.e. a transition from an insulating to a metallic phase induced by disorder in a lattice where all bands are flat. This phenomenon was predicted more than ten years ago for a three-dimensional diamond lattice with flat bands ~\cite{Inverse_prl}, however its experimental observation has remained elusive so far, mainly because of the impracticality of engineering and controlling special three-dimensional lattice geometries. In low-dimensional flat band systems, inverse Anderson transition is rather generally prevented, and a competition between  geometric frustration and Anderson localization is typically observed \cite{Disorder_pra}. However, it has been recently predicted that under certain correlated disorder inverse Anderson transition could be observed in quasi one-dimensional (1D) AB caging systems ~\cite{Inverse_ol}. 

 In this Letter, we experimentally realize a quasi 1D  rhombic chain with synthetic magnetic flux and demonstrate flat-band localization in the momentum-space lattice of ultracold atoms ~\cite{2016_pra,2018_science,2019_NJP,2019_npj}. Thanks to the flexible engineering abilities offered by the artificial lattice in momentum space, we can control on-site correlated disorder in the system and  explore the interplay between disorder and localization, with the observation of the inverse Anderson transition under anti-symmetric binary disorder. Since our artificial lattice platform  is accomplished under the many-body environment ~\cite{Xie2020, An2021a, Xiao2021, Wang2022}, it could offer the possibility to experimentally explore other exotic properties of flat band systems with strongly-correlated particles \cite{Vidal2000,FlachReview2018,Lewenstein2017,Flach2020,Platero2020,Kuno2020}.%

{\bf Theoretical model and experimental realization.} Theoretically, we consider a quasi-1D rhombic lattice with three coupled sublattices (denoted as A, B, and C) as shown in Fig. 1(a). %Under the tight-binding approximation~\cite{2015_pra,2016_pra}, the whole rhombic lattice 
The system is described by the following effective Hamiltonian
\begin{equation}\label{eq1}
\begin{aligned}
&H_\text{eff}=\sum_{n}[-J(e^{i\theta_{3}}\hat{a}_{n}^{\dagger}\hat{b}_{n-1}+e^{-i\theta_{4}}\hat{a}_{n}^{\dagger}\hat{b}_{n}+e^{-i\theta_{2}}\hat{a}_{n}^{\dagger}\hat{c}_{n-1}+\\ & \ \ \ e^{i\theta_{1}}\hat{a}_{n}^{\dagger}\hat{c}_{n}+\text{H.c.})-(\Delta_{n}^{(a)} \hat{a}_{n}^{\dagger} \hat{a}_{n}+\Delta_{n}^{(b)} \hat{b}_{n}^{\dagger}\hat{b}_{n}+\Delta_{n}^{(c)} \hat{c}_{n}^{\dagger}\hat{c}_{n})]
\end{aligned}
\end{equation}  
where $J$ is the coupling amplitude between neighboring sites,  $\hat{a}_{n}, \hat{b}_{n}$ and $\hat{c}_{n}$ ($\hat{a}_{n}^{\dagger}, \hat{a}_{n}^{\dagger}$ and $\hat{a}_{n}^{\dagger}$) are the annihilation (creation) operators for particles at the A, B and C sites of the $n$-th unit cell, $\phi $ is the synthetic magnetic flux of each plaquette defined as $\phi=\sum_{m=i}^{4} \theta _{m}$, and $\Delta^{(i)}_n \; (i=a,b,c)$ are the on-site energy potentials in the three sites at the $i$-th unit cell.

In a clean lattice system, where the on-site energy shifts are set to be zero, $i.e. ~\Delta_n^{(i)}=0$, there are three bands with dispersion relations: $E_{0}=0$ and $E_{\pm }=\pm 2J\sqrt{1+\cos(\phi /2)\cos(k+\phi /2)}$, where $k \; (-\pi \leq k< \pi )$ is the Bloch wavenumber; see  the middle panel of Fig. 1(b). Notably, for $\phi =\pi$ the lattice displays three flat bulk bands [left inset of Fig. 1(b)] with compact localized states. In this case transport is forbidden owing to geometric localization (AB caging effect). Conversely. for $\phi \neq \pi$ the system supports only one flat bulk band and two dispersive bulk bands, as shown in the right inset of Fig. 1(b) for $\phi=0$.

Besides the clean lattice configuration, we can add the on-site static disorder terms to explore the insulating and metallic phases in the rhombic chain. For $\phi \neq \pi$, disorder can induce Anderson localization of dispersive bands and prohibit transport. However, for $\phi=\pi$, i.e. in the fully flat band case, an inverse Anderson transition  can arise: adding disorder breaks the geometric localization of compact states  and enables transport in the lattice \cite{Inverse_ol}.

Experimentally, we engineer a chain of  Aharonov-Bohm rhombic rings along the momentum lattice in a $^{87}\textrm{Rb}$ Bose-Einstein condensate (BEC)  with number  $~\sim2\times10^{5}$~\cite{2020_prl,2018_JOSAB,2021_npj}. As illustrated in Fig. 1(a), the three sublattice sites are experimentally realized in the synthetic dimensions of atomic internal states, $i.e.$ the ground-state hyperfine state $\left| F=1,m_{F}=0 \right>$  for sites A and B, and  $\left| F=2,m_{F}=0 \right>$  for site C. The adjacent sites between different internal states are coupled by Raman transitions, and the adjacent sites in $ F=1$ are coupled via Bragg transitions \cite{NHSE,supp}. The tunneling amplitude is set to be {$J=0.95(5)$} kHz for all the experiments, and the synthetic magnetic flux (denoted as $\phi$) is controlled by the relative phases of coupling lasers \cite{supp}. In total, we implement a chain of  10 unit cells  with open boundary condition (OBC) and mainly focus on the bulk state dynamics.

%\iffalse
\begin{figure}[!t]
	\centering
	\includegraphics[width=1.0\linewidth]{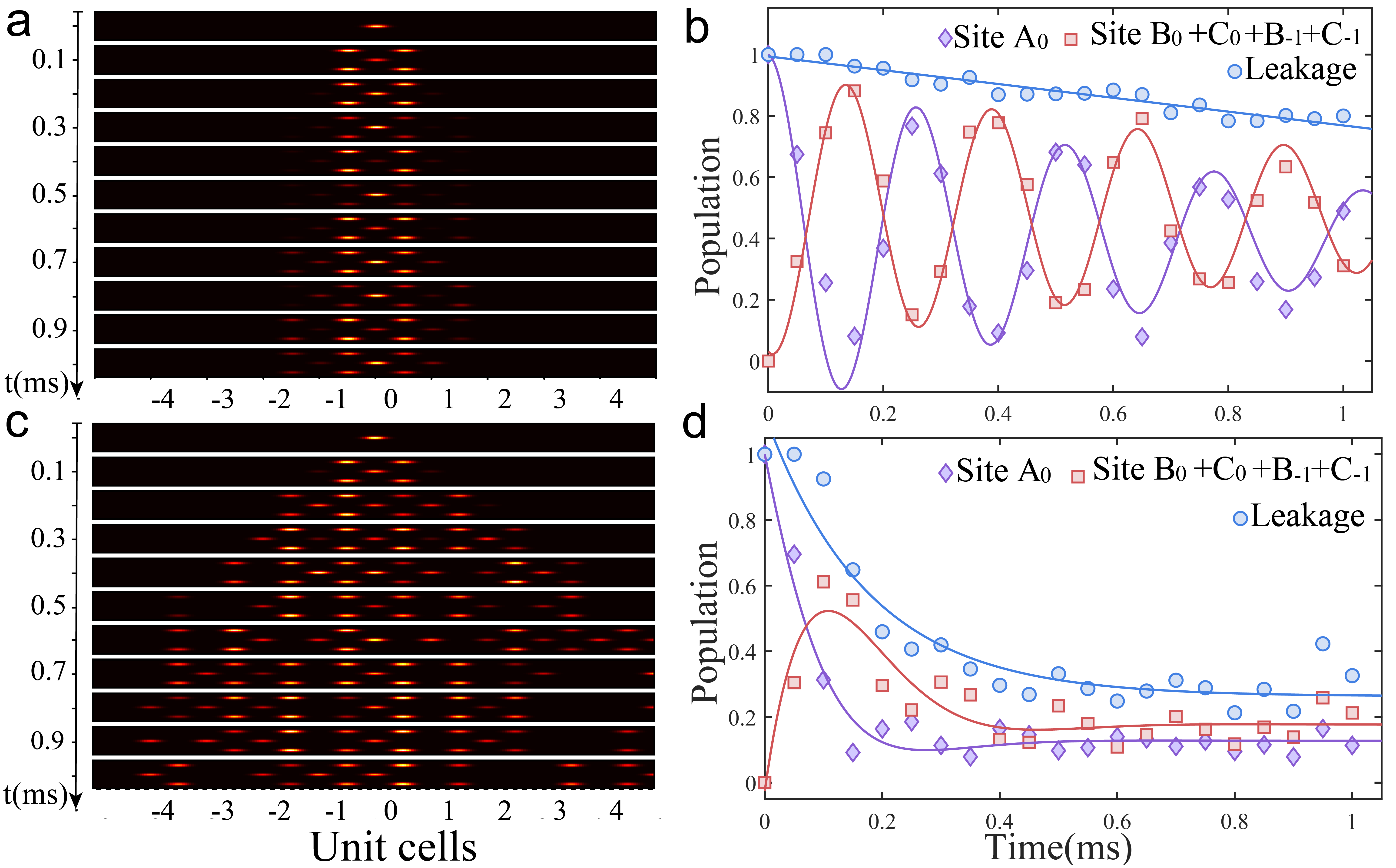}
	\caption{\textbf{Aharonov-Bohm cage dynamics.} \textbf{a.} The Aharonov-Bohm cage evolution with flux $\phi = \pi$ in each plaquette. \textbf{b.} Measuring the population distribution of sites A, B, and C in 1 $ms$. The AB cage effect appears in the middle adjacent unit cells. The small population leakage of the middle cage can be mainly ascribed to the imperfect experimental parameters. The solid lines represent the fitting of experimental data. \textbf{c.} The evolution dynamics with flux $\phi = 0$ in each plaquette. \textbf{d.} The population distribution of sites A, B, and C in the middle unit cell with flux $\phi = 0$. All the fitted solid lines show a fast exponential decay feature . In (a) and (c), the population of all lattice sites has been normalized in each displayed time instant.} 
        \label{f2}
\end{figure}

%\fi

{\bf Aharonov-Bohm cage dynamics.} As Fig. 1(b) shows, the energy spectrum of AB caging limit ($\phi=\pi$ and without on-site energy shift) has three non-dispersive bulk bands. The zero eigenvalue ($E_{0}=0$) has the localized eigenstate $\left| \psi_{n,0}^{bulk} \right>=(b^{\dagger}_{n-1}+c^{\dagger}_{n-1}+b^{\dagger}_{n}-c^{\dagger}_{n})\left| 0 \right>$, in which the $A$ state is not involved. And the other two eigenstates of flat bands are $\left| \psi_{n,\pm}^{bulk} \right>=(b^{\dagger}_{n-1}+c^{\dagger}_{n-1}\mp 2a^{\dagger}_{n}-b^{\dagger}_{n}-c^{\dagger}_{n})\left| 0 \right>$ ~\cite{ABcage_prl,Inverse_ol}. Therefore, the initial excitation at the site $A_n$ can be viewed as the superposition of $\left| \psi_{n,\pm}^{bulk} \right>$, whose dynamics shows the feature of breathing motion and the oscillation frequency is determined by their eigenfrequency difference. In the experiment, we prepare the initial BEC wave packet at site A of the middle unit cell (labeled as $0_{th}$) and observe the evolution process of its population. Figure 2(a) shows the AB caging evolution dynamics in a period of 1 $ms$, which displays the localization and breathing features. As Fig. 2(b) shows, the oscillated population of site A (or B and C) gives the frequency $\omega=$3.72(5) kHz, which is consistent with the energy spectrum of Fig. 1(b) ($\omega=4J$ in theory). {Due to the decoherence of our Raman coupling lattice, the fitted decay lifetime of cage dynamics is 0.70(2) $ms$, which is depicted by the cosine decay of Fig. 2(b). The ideal AB cage sustains compact localized states, while the actual experimental result suffers for a small population leakage as depicted in Fig. 2(b), which is mainly ascribable to the imperfect symmetric neighboring coupling and decoherence~\cite{supp}}. By comparison, we also show the evolution dynamics with $\phi=0$ in Figs. 2(c) and 2(d). The geometric localization feature disappears and each site population of the $0_{th}$ unit cell displays a fast near-exponential decay behavior.

{\bf Inverse Anderson transition.} As mentioned before, the system with non-$\pi$ synthetic magnetic flux possesses one flat band ($E_{0}$) and two dispersive bands ($E_{\pm }$). For $\phi \neq \pi$, the addition of static disorder will generally lift the flat band eigenvalue degeneracy and cause mixing inside this band, which results in diffusion as those states with dispersive bands. 
For $\phi=\pi$, i.e. in the fully flat band case, the addition of uncorrelated on-site static disorder is not able to induce an inverse Anderson transition ~\cite{Disorder_pra}. In fact, for strong uncorrelated disorder ($i.e.$ larger than $2J$) localization features as in the ordinary Anderson localization are observed: the energy bands are mixed and the localization length decreases as the disorder strength increases ~\cite{Disorder_pra}. On the other hand, for weak disorder ($i.e.$ smaller than $2J$) the Anderson localization effect is overwhelmed: the eigenstates from each band are separated by gaps and their localization lengths do not depend on the disorder strength ~\cite{Disorder_pra}.

\begin{figure}[!t]
\centering
\includegraphics[width=1\linewidth]{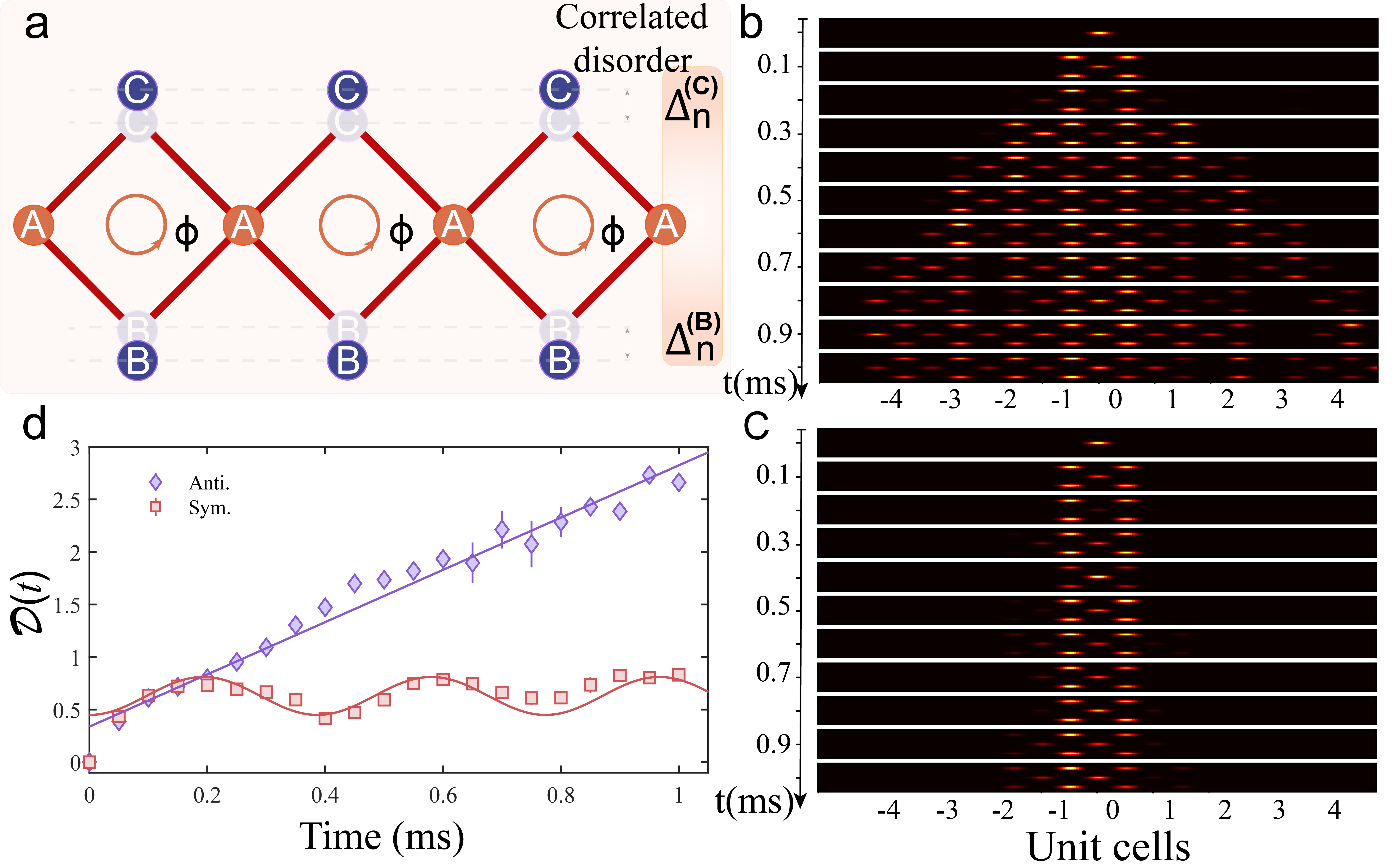}
\caption{\textbf{The localization and transport with symmetric vs. anti-symmetric correlated disorder.} \textbf{a.} Schematic diagram of adding on-site disorders. \textbf{b.} The momentum lattices evolution of ballistic transport with the anti-symmetric on-site disorder. \textbf{c.} The momentum lattices evolution of AB cage localization with the symmetric on-site disorder. \textbf{d.} The compared evolutions of distance $\mathcal{D}(t)$ with a different type of on-site disorder. All above results are set with flux $\phi = \pi$. In (b) and (c), the population has also been normalized in each displayed moment. The above experimental results refer to a single realization of correlated Bernoulli disorder.}
\label{f2}
\end{figure}

In contrast, the correlated on-site static disorder in sublattices B and C can induce transport, i.e. an inverse Anderson transition can be observed for correlated disorder in the system ~\cite{Inverse_ol}. There are two typical cases of correlated disorder, $i.e.$ the symmetric ($\Delta_{n}^{(b)}=\Delta_{n}^{(c)}$) and anti-symmetric  ($\Delta_{n}^{(b)}=-\Delta_{n}^{(c)}$) correlated disorder, where $\Delta_{n}^{(b)}$ and $\Delta_{n}^{(c)}$ are independent stochastic variables with the same probability density function of zero mean. As shown in~\cite{Inverse_ol} and briefly reviewed in ~\cite{supp}, the symmetric case shares similar localization dynamics as the AB caging:  band degeneracy is lifted by the disorder but the eigenstates are still compact localized states. More striking behaviors appear in the anti-symmetric correlated scenario $\Delta_{n}^{(b)}=-\Delta_{n}^{(c)}$, where the existence of extended states can be observed, indicating the disorder-induced transport regime (inverse Anderson transition). The localization or mobility features dynamics for anti-symmetric correlated disorder depends on the forms of the probability density function of the stochastic variables $\Delta_{n}^{(b)}=-\Delta_{n}^{(c)}$~\cite{Inverse_ol}. Specifically. the uniform distribution of probability density function prevents mobility, while the Bernoulli distribution, where $ \Delta^{(b)}_n$ can take randomly only the two values $\pm \Delta$,  induces ballistic transport.

\begin{figure}[!t]
\centering
\includegraphics[width=0.85\linewidth]{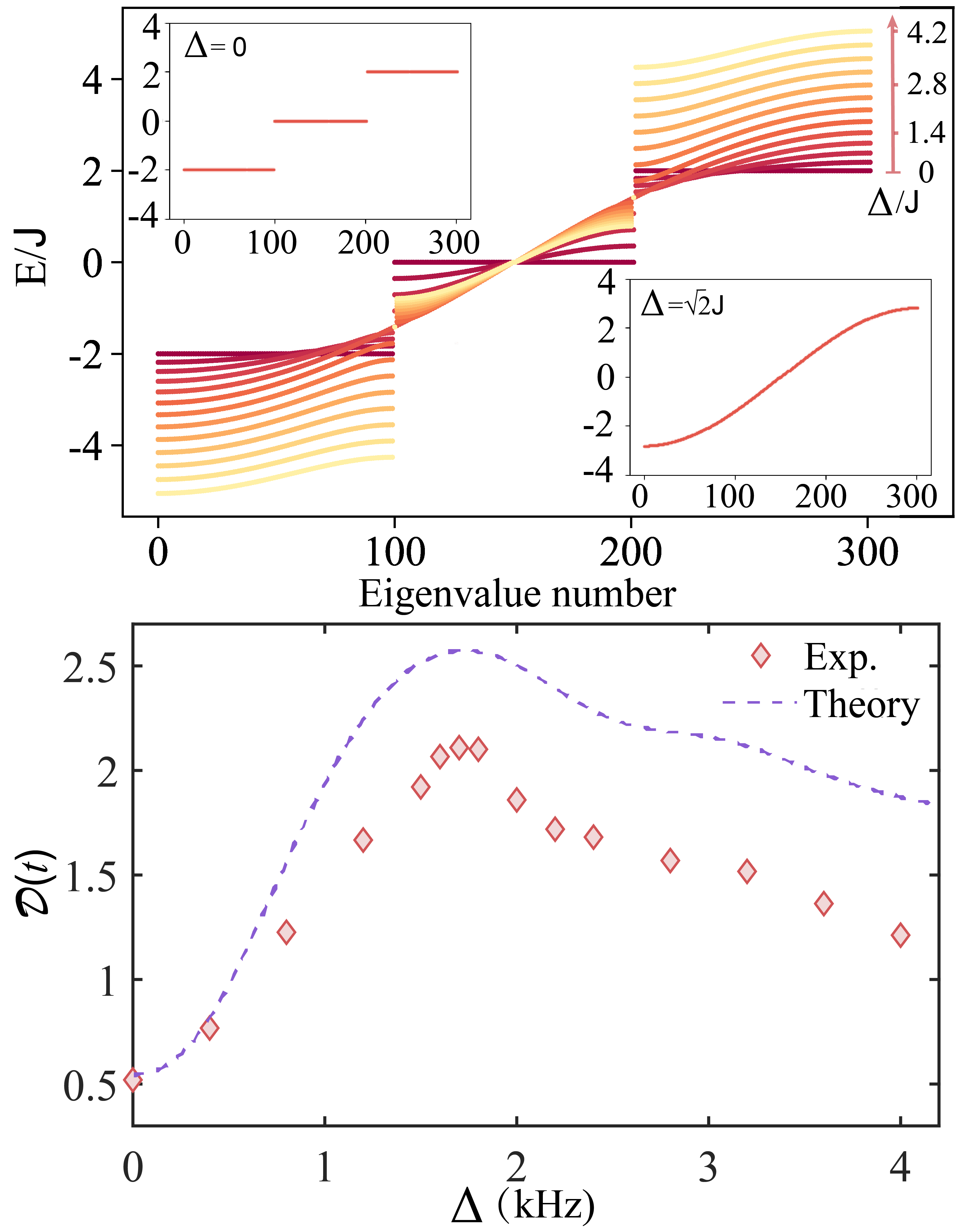}
\caption{\textbf{The energy bands and evolution with anti-symmetric Bernoulli disorder.} \textbf{a.} The quasienergy spectrum for different on-site disorder strengths, where we have calculated eigenvalues of a chain with 100 unit cells. The left and right inset correspond to the smallest and largest bandwidth, respectively. \textbf{b.} The measured distance with various disorder strengths in $0.6~ms$. The inset  line corresponds to the theoretical calculated distance. The above measurements refer to a single realization of Bernoulli disorder.}
\label{f4}
\end{figure}

To characterize the transport dynamics, we excite the system in a single site and monitor the spreading dynamics using the second-order moment of position operator $\mathcal{D}^2(t)$, given by
\begin{equation}\label{eq1}
\mathcal{D}^{2}(t)=\sum_{n}n^{2}(\left| a_{n}\right|^{2}+\left| b_{n}\right|^{2}+\left| c_{n}\right|^{2}).
\end{equation} 
A distance $\mathcal{D}(t)$ that linearly  grows with time, i.e. $\mathcal{D}(t)\sim t$ , is the clear signature of ballistic transport regime. Figure 3(a) shows a schematic of the setup where on-site disorder energy shift is added in our momentum lattice system~\cite{supp}, whereas the experimental results of wave packet spreading are illustrated in Figs.3(b-d). In the experiment, the initial excitation of BEC wavepacket is in the sublattice A of unit cell at site $n = 0$, $i.e. ~a_{n}(0) = \delta _{n,0}$ and $b_{n}(0) = c_{n}(0) = 0$.  The figures clearly show that the symmetric and anti-symmetric correlated disorder display completely different evolution dynamics: while ballistic transport is clearly observed in the antisymmetric case, corresponding to an inverse Anderson transition, wave spreading is prevented in the symmetric case [Fig.3(d)].

To further unveil the features of the anti-symmetric disorder case, in Fig.4(a) we show the energy spectrum of the finite rhombic chain with the flux $\phi = \pi $ under antisymmetric disorder with a Bernoulli distribution, parametrized in the  strength $\Delta$ of disorder. We can see that the energy spectrum changes from the flat bands at $\Delta=0$ to dispersive bands as the disorder strength increase from zero to a large value. The theoretical analysis~\cite{supp} indicates that for $\Delta>0$ the Hamiltonian displays absolutely continuous spectrum with three dispersive  Bloch bands, and their gaps vanish at $\Delta=\Delta_{0} = \sqrt{2}J$. At this turning point, the transport is fastest due to the largest bandwidth. Experimentally, we fix the evolution time to be $t= 0.6 \;ms$ and measure the distance $\mathcal{D}(t)$ for different disorder strengths. As shown in Fig. 4(b), our experimental results support this prediction. However, due to the finite size effect, our experimentally measured turning amplitude is $\Delta_0=1.7 J$, which is larger than the theoretically predicted value. If the rhombic chain is long enough or the measured evolution time is much longer, our numerical results show that the turning point is approaching the prediction of the ideal infinite chain \cite{supp}. 
The aberration of the dynamics at larger disorder strength is mainly due to the decoherence of our system, which slower the quantum transport and makes the distance smaller. This effect is more important for fast transport, thus the distance is reduced more for larger disorder strength.

{\bf Conclusion.} In conclusion, we reported on the experimental realization of a quasi-1D rhombic chain in a momentum lattice of ultracold atoms, and demonstrated Aharonov-Bohm caging dynamics when a synthetic gauge field is applied. Thanks to the flexible engineering abilities provided by our platform, we could finely tune the tunneling process and introduce controllable disorder into the system, unraveling the interplay between disorder and transport in a system with geometric localization. In particular, we reported on the first experimental observation of an inverse Anderson localization in a flat band system ~\cite{Inverse_prl}, i.e. the delocalization of the wave functions and ballistic transport induced by suitable correlated static disorder in the AB system. The AB photonic cage and inverse Anderson transition reported in our experiments are observed in a regime where particle interaction is negligible, however thanks to the many-body environment offered by the ultracold gas platform our synthetic lattice setup could be feasible for the experimental demonstrations of other exotic effects arising in strongly-correlated flat bands systems.

\begin{acknowledgments}
{\it Acknowledgement:-}
We acknowledge the support from the National Key Research and Development Program of China under Grant No. 2018YFA0307200, the National Natural Science Foundation of China under Grants No. U21A20437 and No. 12074337, Natural Science Foundation of Zhejiang Province under Grant No. LR21A040002, Zhejiang Province Plan for Science and Technology Grant No. 2020C01019, and the Fundamental Research Funds for the Central Universities under Grant No. 2021FZZX001-02.
\end{acknowledgments}

\bibliographystyle{apsrev4-2}
\bibliography{ABcage}

\end{document}